\documentclass[aps,superscriptaddress,floatfix]{revtex4}
\usepackage{graphicx}
\usepackage{amssymb,amsfonts,amsmath,MnSymbol}
\usepackage[usenames,dvipsnames]{pstricks}
\pdfoutput=1
\newcommand{\prs}[1]{{\left(#1\right)}}
\newcommand{\prob}[1]{{\mathcal{P}\prs{#1}}}
\newcommand{\ie}{\textit{i.e.\ }}

\begin{document}

\title{The interplay of microscopic and mesoscopic structure in complex networks}

\author{J\"org Reichardt}
\affiliation{Institute for Theoretical Physics, University of W\"{u}rzburg,
W\"{u}rzburg, Germany}
\affiliation{Complexity Sciences Center, UC Davis, Davis, USA}
\author{Roberto Alamino}
\author{David Saad}
\affiliation{The Nonlinearity and Complexity Research Group, Aston University, Birmingham,UK},

\begin{abstract} % no more than 150 words, including "here we show"
\noindent  Not all nodes in a network are created equal. Differences and similarities exist at both individual node and group levels. Disentangling single node from group properties is crucial for network modeling and structural inference. Based on unbiased generative probabilistic exponential random graph models and employing distributive message passing techniques, we present an efficient algorithm that allows one to separate the contributions of individual nodes and groups of nodes to the network structure. This leads to improved detection accuracy of latent class structure in real world data sets compared to models that focus on group structure alone. Furthermore, the inclusion of hitherto neglected group specific effects in models used to assess the statistical significance of small subgraph (motif) distributions in networks may be sufficient to explain most of the observed statistics. We show the predictive power of such generative models in forecasting putative gene-disease associations in the Online Mendelian Inheritance in Man (OMIM) database. The approach is suitable for both directed and undirected uni-partite as well as for bipartite networks.
 \end{abstract}

\maketitle 

%From Nature's Guide to Authors
%Contributions should be organized in the sequence: title, text, methods, references, Supplementary Information line (if any), acknowledgements, author contributions, author information (containing data deposition statement, competing interest declaration and corresponding author line), tables, figure legends.

%A typical Article contains about 3,000 words of text and, additionally, five small display items (figures and/or tables) with brief legends, reference list and methods section if applicable.
%
%%%%%%%%%%%%%%%%%%%%%%%%%%%%%%%%%%%%%%%%%%%%%%%%%%%%%%%%%%%%%%%%
\begin{small}
%Introduction, no more than 500 words
Networks are fascinating objects. Charting the interactions between system constituents, abstracted as edges and nodes, has allowed us to marvel the interconnectedness of systems and appreciate their complexity. Whether in understanding foodwebs~\cite{TopologyFoodweb}, social communities~\cite{Girvan}, protein-interaction~\cite{FuncPredReview,Pinkert}, metabolism~\cite{GuimeraMeta}, neural networks~\cite{Honey} or communication~\cite{BarabasiAlbert} the network-metaphor has been highly successful in advancing our understanding of complex systems. While being esthetic charts of complex systems, networks only reveal insights through rigorous statistical analysis and modeling.
%Networks as Abstractions - Node property / function links
The abstraction of complex systems as networks, \ie graphs connecting nodes through links, provides a tremendous simplification. Not all network constituents are created equal; network structure strongly depends on node properties and functions which may differ radically among individual nodes, or may be shared by a whole group of nodes. However, in many circumstances, node properties or functions may not be accurately or not completely known in contrast to the interactions between nodes, which often can be mapped accurately and efficiently. A primary goal of network research is to deduce unobserved, or latent, node properties through structural analysis.

%structural analysis descriptive vs. probabilistic modelling
Methodologically, structural analysis can proceed using either purely descriptive statistics or a generative model coupled with inference techniques; both of which are complementary. However, a descriptive approach can only make assertions about a single network, while generative models gives rise to a whole ensemble of networks that are statistically equivalent to the observed dataset. Such ensembles are vital to the study of dynamical processes on networks but above all, they allow us to potentially differentiate between more and less important structural features. We will hence use generative models in our approach.

Thematically, research on network structural features has two foci. One is the study of \emph{microscopic} properties, such as very small subgraphs of a given size known as motifs~\cite{AlonMotifs} and their respective distributions in networks, or the properties of single nodes, like node degree, centrality, betweenness or page-rank, which are correlated with latent node characteristics such as expansiveness, esteem, functional importance or authority. The second investigates \emph{mesoscopic} structural features indicative of properties or functions \emph{shared} by groups of nodes. Differing link densities between entire classes of nodes, assortative or dissortative mixing~\cite{NewmanMixing}, community structure~\cite{SantoReview,ReichardtPRL} or more generally, block structures~\cite{DoreianBook,ReichardtWhite}, are all examples of the latter. Interestingly, the two themes run in parallel. When modeling degree distributions or analyzing motif distributions, group effects are rarely taken into account, while individual node properties are generally neglected in inferring latent node classes from network structure via community or block structure detection algorithms.

With this paper, we intend to close this gap. We present a principled probabilistic approach to the inference of latent node classes based on a generative model that includes node specific features; these provide a more realistic model and enable one to quantify probabilistic measures and confidence levels of the observed structural details. To estimate model parameters, we employ distributive message-passing techniques, with computational complexity scaling linearly with the problem size. Using real world data, social and biological networks, we demonstrate the significance of node specific parameters to the quality of latent class inference. Through an example from neuroscience, we show that including group specific effects in the random null model used to assess the statistical significance of motif counts may provide a group based explanation for many motifs. Finally, we demonstrate how such generative probabilistic models can be used for the prediction of putative new gene-disease associations.

%
% FIG 1 ----------
\begin{figure*}
\includegraphics{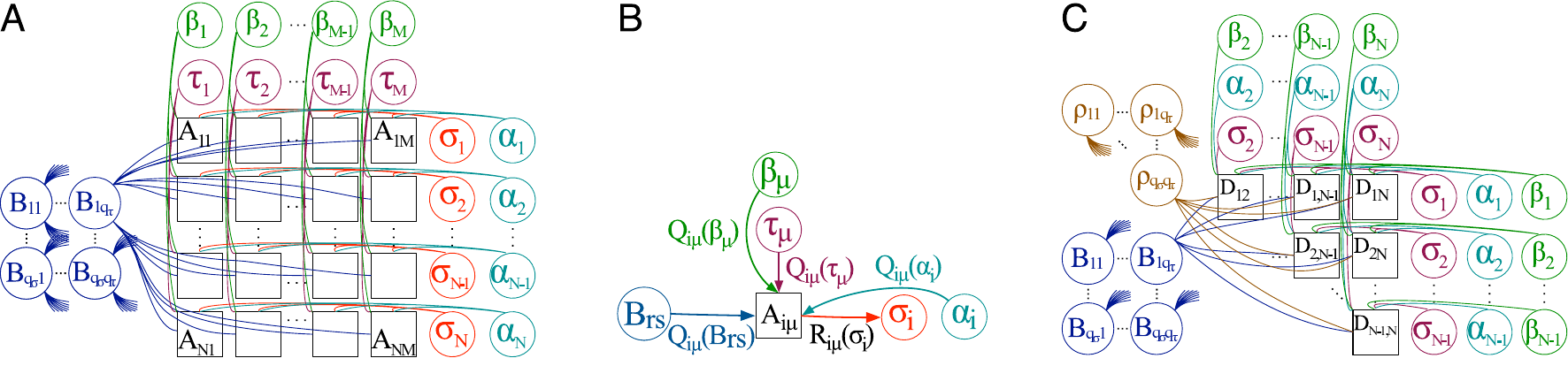}
\caption{Factor graphs and an  example of an elementary message passing update. Factors of the likelihood function are represented as squares, variables of the generative model as circles. Connections indicate which variables enter the calculation of which factor. \textbf{A)} For a bipartite actor-event networks represented by an $N\times M$ adjacency matrix $A_{i\mu}$, class label $\sigma_i$ and activity $\alpha_i$ of actor $i$ enter in the calculation of all factors in row $i$. Equivalently, class label $\tau_\mu$ and popularity $\beta_\mu$ of event $\mu$ enter in the calculation of all factors in column $\mu$. The variables $B_{rs}$ denoting preference of actors in class $r$ for events in class $s$ enter in every factor. Note that while each factor depends on only $\mathcal{O}(1)$ variables, the $\sigma$ and $\alpha$ variables enter in the calculation of $\mathcal{O}(N)$, the $\tau$ and $\beta$ variables in $\mathcal{O}(M)$ and the $B_{rs}$ variables in $\mathcal{O}(NM)$ factors. \textbf{B)} Pictorial representation of the messages involved in calculating $R_{i\mu}(\sigma_i)$ sent from factor $A_{i\mu}$ to variable $\sigma_i$ according to equation~(\ref{SpecialUpdate}). \textbf{C)} For directed networks represented by non-symmetric $N\times N$ adjacency matrices, the factors correspond to dyads $D_{ij}=(A_{ij},A_{j i})$. Additional to the interclass preference matrix, a symmetric matrix of reciprocities $\rho_{rs}$ enters into the model (\ref{ERGM3}). Every node $i$ carries a single class label $\sigma_i$, activity $\alpha_i$ and attractiveness parameter $\beta_i$. The variables associated with node $i$ enter in the calculation of factors in both row $i$ and column $i$.
}\label{DepGraph}
\end{figure*}
%------------

%--------------------------
\section{Methods}
\subsection{Exponential random graphs}
When inferring structural features in networks with probabilistic generative models, we have to establish what constitutes a good model. Within the infinite set of possible models, exponential random graph models (ERGMs)~\cite{HollandLeinhardt,WassermanPattison} stand out as combining desired properties that make them the preferred choice for the task. ERGMs are mean unbiased and make the observed data maximally likely. They are maximum entropy models thus ensuring that one avoids adding redundant assumptions to the network's given structural features. In other words, they parameterize the largest ensemble of networks compatible with our observations, while making the observed network typical for the ensemble. Plus, their parameters have a very intuitive interpretation.

Within the framework of ERGMs it is easy to combine node specific with group specific effects. Consider a given, bipartite network specified by its $N\times M$ adjacency matrix $\mathbf{A}$, representing for instance the attendance of $N$ actors in $M$ events. If actor $i$ has attended event $\mu$, then $A_{i\mu}=1$ and otherwise $A_{i\mu}=0$. Equally, $\mathbf{A}$ could represent the choices of $N$ consumers from a list of $M$ products, or the association of $N$ diseases with $M$ different genes. The possibilities are many and we will use the actor-event picture, but without limiting the applicability of the model to this case alone.

We restrict ourselves to \emph{dyadic} models, \ie we assume the entries of the adjacency matrix $A_{i\mu}$ to be modeled by the conditionally independent random variables $D_{i\mu}\in \{0,1\}$.
A simple ERGM that captures both individual (actor- and event-specific) and  group-specific factors is given in terms of the odds ratio of actor $i$ attending event $\mu$~\footnote{The reader wondering where the ``exponential'' notion is in our specification of ERGMs may rewrite the product of parameters in equation (\ref{ERGM1}) as $\exp(a_i + b_\mu + C_{\sigma_i\tau_\mu})$ where $a_i=\ln \alpha_i/(1-\alpha_i)$, $b_\mu=\ln \beta_\mu/(1-\beta_\mu)$, and $C_{\sigma_i\tau_\mu}=\ln  B_{\sigma_i\tau_\mu}/(1-B_{\sigma_i\tau_\mu})$}:
\begin{equation}
\frac{\prob{D_{i\mu}=1|\vec{\theta}}}{\prob{D_{i\mu}=0|\vec{\theta}}}=\frac{\alpha_i}{(1-\alpha_i)}\frac{\beta_\mu}{(1-\beta_\mu)}\frac{B_{\sigma_i\tau_\mu}}{(1-B_{\sigma_i\tau_\mu})}.
\label{ERGM1}
\end{equation}
The shorthand $\vec{\theta}$ denotes the set of all model parameters, \ie in this case $\vec{\theta}\equiv(\alpha_1,..,\alpha_N,\beta_1,..,\beta_M,\sigma_1,..,\sigma_N,\tau_1,..,\tau_M, \mathbf{B})$. Of these, only a small subset is relevant for an individual dyad $D_{i\mu}$. The parameter $\alpha_i \in (0,1)$ denotes the global \textit{activity} of actor $i$, higher $\alpha_i$ meaning higher odds of attending any event. Correspondingly, $\beta_\mu \in (0,1)$ denotes the global \textit{popularity} of event $\mu$. Furthermore, every actor $i$ and every event $\mu$ carry a class index $\sigma_i$ and $\tau_\mu$, respectively\footnote{The number of classes is determined a priori here. It represents a free parameter that defines the coarseness or resolution of the grouping sought.}.
The matrix $B_{rs} \in (0,1)\forall\mbox{ } r,s$, models the data at a coarser, group specific level, denoting the \textit{tendency} or \textit{preference} of an actor of class $r$ to attend an event of class $s$. Higher entries mean higher odds for the attendance of any actor of class $r$ to any event of class $s$. The matrix $B_{rs}$ is also called a block model of the data.

The rich literature on ERGMs \cite{SpecIssueSoc} has generally assumed prior knowledge of the class labels $\sigma_i$ and $\tau_\mu$ in (\ref{ERGM1}), or other covariates \cite{FienbergWasserman,HollandLaskeyLeinhardt,WangWong,BianconiNodeFeatures}. Then, learning the parameters of (\ref{ERGM1}) practically reduces to a simple logistic regression. However, the learning task is considerably more complicated if the latent class labels $\sigma_i$ and $\tau_\mu$ are unknown and need to be inferred as hidden variables. On the other hand, a growing body of work is dedicated to the development of efficient algorithms for learning general stochastic block models~\cite{NowickiSnijders,SnijdersBlocks,Daudin,GuimeraMissing,NonparamModularity} including the hidden assignment of nodes into classes, but without the incorporation of node specific effects, \ie a model specified by
\begin{equation}
\frac{\prob{D_{i\mu}=1|\vec{\theta}}}{\prob{D_{i\mu}=0|\vec{\theta}}}=\frac{B_{\sigma_i\tau_\mu}}{1-B_{\sigma_i\tau_\mu}}.
\label{ERGM2}
\end{equation}
This model is also referred to, with slight variations, as infinite relational model~\cite{Kemp} or mixed membership stochastic block model~\cite{Airoldi2}. The common thread among these is the goal to model network structure entirely in terms of groups of nodes. Some attempts to include the estimation of node specific effects have resulted in biased models~\cite{Morup,LeichtMixture}. Within the framework of ERGMs, node and group specific properties have been combined in so called latent space models~\cite{HoffLatentSpace,Krivitsky} where nodes are assigned a position in an abstract space and links form as a function of their distance. Such models are well motivated for social networks, where homophily is a central mechanism of link formation and proximity in the latent space may be interpreted as similarity. Yet they are less general than stochastic block models being caught in the predicament of placing groups of nodes with similar interaction partners in close proximity while at the same time having to place them apart if these nodes are not densely connected among each other. Our approach facilitates parameter estimates and latent class inference in model (\ref{ERGM1}) which combines node specific effects with the more general stochastic block models for group structure.

\subsection{Model Inference}
To describe the algorithm for estimating the parameters $\vec{\theta}$, we first write the likelihood of the entire observed network adjacency matrix $\mathbf{A}$ in terms of our model (\ref{ERGM1}):
\begin{equation}
\mathcal{L(\vec{\theta})}\equiv\prob{\mathbf{A}|\vec{\theta}}=\prod_{i\mu}\prob{D_{i\mu}=A_{i\mu}|\vec{\theta}}
\label{LL}
\end{equation}
For a dyadic model, the likelihood factorizes into terms that involve parameters associated with only two nodes.

Commonly used methods to estimate the parameters and hidden variables in such a model are to employ maximum likelihood (ML) techniques in the form of an expectation-maximization type algorithm or Monte Carlo sampling \cite{statnet}. We prefer a Bayesian approach, based on Maximum A Posteriori (MAP) estimates that does not incur the computational cost of Monte Carlo sampling while being less sensitive to initial conditions and  more stable numerically than ML, especially as the parameters which maximize (\ref{LL}) may lie on the the borders of the admissible interval $(0,1)$. Furthermore, the MAP approach provides a natural Occam's razor as the posterior distributions of parameter estimates can only reduce in variance with the provision of more data, while the ML approach assumes point estimates or $\delta-$functions for the posterior from the start. This is an important feature of the Bayesian approach as it provides a natural limit for the number of inferred classes and confidence levels in the assignments. Classes cannot be arbitrarily small if the posterior for the inter-class link preference $\mathbf{B}$ is to be localized. In contrast, under an ML approach the likelihood increases monotonically when more and hence smaller classes are used and model selection criteria, as in~\cite{BianconiNodeFeatures}, are needed. Finally, Bayesian techniques offer a principled way to incorporate prior domain knowledge for obtaining a more accurate approximate marginal posterior distribution $\prob{\theta_k|\mathbf{A}}$, where $\theta_k$ represents one of the parameters $\alpha_i,\sigma_i,\beta_\mu,\tau_\mu$ or $B_{rs}$.

A \textit{message passing} or belief propagation algorithm provides a principled way to calculate approximate posterior marginal distributions \cite{MacKay,AdMeanField}. The starting point for this algorithm is a so-called \textit{factor-} or \textit{dependency-graph}, a graphical representation of the probabilistic dependencies between the variables (model parameters) we wish to infer from the data, and the individual factors that constitute the likelihood (\ref{LL}). Figure \ref{DepGraph}A shows this for the case of a bi-partite network, likelihood (\ref{LL}) and model (\ref{ERGM1}).

The algorithm proceeds by exchanging messages, conditional probabilities, between factors and variables connected in the dependency graph until convergence. Using the definitions:
\begin{eqnarray}
R_{i\mu}(\theta_k)& \equiv &\prob{D_{i\mu}=A_{i\mu}|\theta_k,\mathbf{A} \backslash A_{i\mu}}\mbox{ and}\nonumber\\
Q_{i\mu}(\theta_k) & \equiv & \prob{\theta_k|\mathbf{A} \backslash A_{i\mu}},
\label{MessageDef}
\end{eqnarray}
one can interpret $R_{i\mu}(\theta_k)$ (\textit{R-Message}) as the likelihood of a single observed matrix entry $A_{i\mu}$ given only the parameter $\theta_k$ and all the data matrix except for entry $A_{i\mu}$. Equally, $Q_{i\mu}(\theta_k)$ (\textit{Q-Message}) is interpreted as the posterior probability distribution of parameter $\theta_k$ given the entire data matrix except for entry $A_{i\mu}$.  For the sake of notational economy, we have adopted to identify functions by their argument. It is to be understood that $R_{i\mu}(\alpha_i)$ is a different function than $R_{i\mu}(\beta_\mu)$ and \emph{not} the same function $R_{i\mu}(x)$ evaluated at the points $\alpha_i$ and $\beta_\mu$ as should be clear from the definitions (\ref{MessageDef}).

Formally, we obtain the R-Message from $A_{i\mu}$ to $\theta_k$, by integrating out all parameters except $\theta_k$ from a likelihood function
\begin{equation}
R_{i\mu}(\theta_k)  =  \sumint \prob{D_{i\mu}=A_{i\mu}|\vec{\theta},\mathbf{A} \backslash A_{i\mu}}\prob{\vec{\theta}\backslash\theta_k|\theta_k,\mathbf{A} \backslash A_{i\mu}}d\vec{\theta}\backslash\theta_k
\end{equation}
Using the independence of given data entries $A_{i\mu}$ we can readily identify $\prob{A_{i\mu}|\vec{\theta},\mathbf{A} \backslash A_{i\mu}}$ with the $\prob{A_{i\mu}|\vec{\theta}}$ of (\ref{ERGM1}). Assuming the joint distribution $\prob{\vec{\theta}|\mathbf{A} \backslash A_{i\mu}}$ factorizes with respect to every single $\theta_k$, one obtains the following closed set of equations:
\begin{eqnarray}
R_{i\mu}(\theta_k=x) & = & \sumint \prob{D_{i\mu}=A_{i\mu}|\vec{\theta}}\prod_{\ell\neq k}Q_{i\mu}(\theta_\ell)d\theta_\ell\mbox{ and} \nonumber\\
Q_{i\mu}(\theta_k=x) & \propto & \prob{\theta_k}\prod_{j\nu\neq i\mu}R_{j\nu}(\theta_k=x).
\label{GeneralUpdate}
\end{eqnarray}
Although the factorization assumption may seem strong, it merely means that the Q-Messages $\prob{\theta_k|\mathbf{A}\backslash A_{i\mu}}$ for any two variables $\theta_k$ and $\theta_\ell$ with $k\neq \ell$ are assumed independent. Given that these distributions are conditioned on the \emph{entire} data matrix except for one entry, the error we make using this assumption is considered negligible for large systems. The form of calculating $Q_{i\mu}(\theta_k=x)$ in (\ref{GeneralUpdate}) follows  directly from Bayes' theorem and $\prob{\theta_k}$ is the distribution we use to include prior information. These equations can be iterated until convergence after which we finally obtain the desired approximate marginal posterior distribution, for every single parameter, as:
\begin{equation}
\prob{\theta_k|\mathbf{A}} \propto \prob{\theta_k}\prod_{i\mu}R_{i\mu}(\theta_k).
\end{equation}

To illustrate these ideas, explicit update equations for the inference of the hidden class index $\sigma_i$ of node $i$ appear below. Expressions for other parameters will be reported elswhere. With
\begin{eqnarray}
X_{rs}^{i\mu} & \equiv &\int \prob{D_{i\mu}=A_{i\mu}|\alpha_i,\beta_\mu,\sigma_i=r,\tau_\mu=s,B_{rs}}\times \nonumber\\
& & Q_{i\mu}(\alpha_i)Q_{i\mu}(\beta_\mu)Q_{i\mu}(B_{rs})d\alpha_id\beta_\mu dB_{rs} ~,
\end{eqnarray}
we can write for the R- and Q-Messages between $A_{i\mu}$ and $\sigma_i$:
\begin{eqnarray}
R_{i\mu}(\sigma_i=r) & = &\sum_{s}X_{rs}^{i\mu}Q_{i\mu}(\tau_\mu=s)\label{Rimur}\mbox{ and}\nonumber\\
Q_{i\mu}(\sigma_i=r) & \propto & \prob{\sigma_i=r}\prod_{\nu\neq \mu}R_{i\nu}(\sigma_i=r).
\label{SpecialUpdate}
\end{eqnarray}
%

%%FIG 2 -------------
\begin{figure*}
\includegraphics[width=18cm]{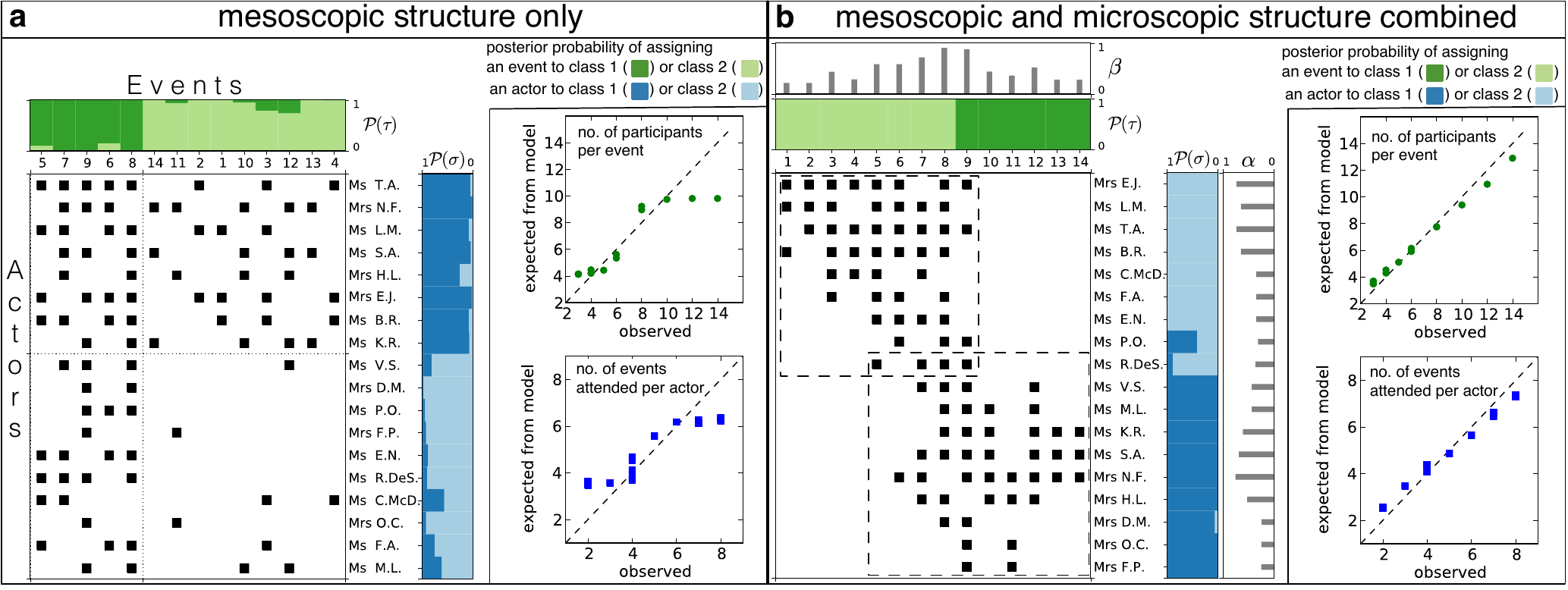}
\caption{Attendance record of 18 women (rows) to 14 informal social events (columns) due to ethnographers Davis, Gardner and Gardner \cite{DeepSouth}. Black squares indicate attendance. \textbf{a)} Attendance matrix with posterior probability of class assignment for actors $\prob{\sigma}$ and events $\prob{\tau}$ as found by learning a standard stochastic block model (\ref{ERGM2}). Classification inferred divides events according to number of attendants and actors according to number of events participated in. The Inset shows the observed numbers of attendances do not agree well with the expectations due to model (\ref{ERGM2}). \textbf{b)} The same attendance matrix as in a) but reordered due to the classification given in the original study indicated by dashed boxes \cite{DeepSouth}. Posterior probability of class assignments inferred using model (\ref{ERGM1}) is almost perfectly compatible with the expert's classification. Including node specific popularity and activity parameters $\beta$ and $\alpha$ allows to match observed numbers of attendances vs.\ expectations from model (\ref{ERGM1}) as shown in inset.}
\label{DavisMatrix}
\end{figure*}
%%------------------
%%
%
The dependency graph greatly facilitates setting-up these update equations. Following the rules that R-Messages are always sent from factors to variables and Q-Messages from variables to factors; and that in R-Messages, we sum or integrate over the incoming Q-messages, while Q-Messages are proportional to the product of incoming R-Messages, we can write the equations based on the dependency graph. Figure~\ref{DepGraph}B shows a detail of~\ref{DepGraph}A focussing on factor $A_{i\mu}$ to illustrate the messages involved in the calculation of $R_{i\mu}(\sigma_i)$ sent to variable $\sigma_i$ as in~(\ref{SpecialUpdate}).
Generalizing the algorithm and update equations to undirected, uni-partite networks is straightforward. A more difficult task is the generalization to directed networks.
Directed uni-partite networks are generally represented by an $N\!\times\! N$ adjacency matrix $\mathbf{A}$. They may exhibit a feature that undirected networks cannot have: non-trivial reciprocity, \ie the fact that the link from node $i$ to node $j$ is generally not independent of the link from node $j$ to node $i$. Still, we can model such networks using dyads $(i,j)$ with $i<j$.  But now, a pair of nodes can not only be connected or disconnected, but we have four different states a dyad can be in: it can be disconnected: $D_{ij}=(0,0)$, with a connection running only from node $i$ to $j$: $D_{ij}=(1,0)$, or only from $j$ to $i$: $D_{ij}=(0,1)$ or the connections between $i$ and $j$ may be reciprocated: $D_{ij}=(1,1)$. Hence, we have to incorporate a parameter that can model this reciprocity. Instead of one global parameter, here we allow reciprocities to differ depending on the latent classes:
\begin{align}
\frac{\prob{D_{ij}=(1,1)|\vec{\theta}}}{\prob{D_{ij}=(0,0)|\vec{\theta}}}=&\frac{\rho_{\sigma_i\sigma_j}}{(1-\rho_{\sigma_i\sigma_j})}\frac{\prob{D_{ij}=1|\vec{\theta}}}{\prob{D_{ij}=0|\vec{\theta}}}\frac{\prob{D_{ji}=1|\vec{\theta}}}{\prob{D_{ji}=0|\vec{\theta}}}\nonumber\\
\frac{\prob{D_{ij}=(1,0)|\vec{\theta}}}{\prob{D_{ij}=(0,0)|\vec{\theta}}}=&\frac{\prob{D_{ij}=1|\vec{\theta}}}{\prob{D_{ij}=0|\vec{\theta}}}
\label{ERGM3}
\end{align}
Larger parameter values $\rho_{rs} \in (0,1)$ increase the odds of finding reciprocated edges between nodes in classes $r$ and $s$. Similar to (\ref{ERGM1}), the parameters $\alpha_i$ and $\beta_i$ model the tendency of node $i$ to initiate and attract links, respectively. The model amounts to coupling two independent models of the type~(\ref{ERGM1}) via the reciprocity parameters.

The likelihood of the observed data under such a model for given parameters is then given analogously to (\ref{LL}) by
\begin{equation}
\mathcal{L}(\vec{\theta})\equiv\prob{\mathbf{A}|\vec{\theta}}=\prod_{i<j}\prob{D_{ij}=(A_{ij},A_{ji})|\vec{\theta}}
\label{LL2}
\end{equation}
and factorizes into terms involving only the parameters of two different nodes. Figure \ref{DepGraph}C shows the dependency graph for such a directed network. With the specification of the model and the dependency graph, the update equations are derived in the same way as above.

%----------------------------
\section{Results}
Next, we will demonstrate the impact of microscopic (node specific) effects on inferred mesoscopic latent class structure by comparing model (\ref{ERGM1}) with the less expressive standard stochastic block model (\ref{ERGM2}). For this, we use a dataset from sociology: the  \textit{Southern Women}. Then, we will show the importance of mesoscopic group effects to the interpretation of microscopic structural features by studying the motif distributions in the neural network of the nematode \textit{C. elegans}. Last, we will determine the influence of both node specific and group specific effects on the accuracy of predicting new links in networks. To this end, we compare model (\ref{ERGM1}) with both a less expressive model and a more expressive model in terms of classification accuracy and predictive power on the network of gene-disease associations from the Online Mendelian Inheritance in Man (OMIM) database.

\emph{Southern women:} This classic bipartite data set is due to ethnographers Davis, Gardner and Gardner~\cite{DeepSouth}. A $18\times 14$ matrix records the attendance of 18 women in southern Alabama to 14 informal social events over the course of a nine month period in the 1930s. The authors' aim was to study how an individual's social class influences her pattern of informal social interaction. Based on intuition and experience in the field, but without formal analysis, the authors suggested the existence of two latent classes of 9 women each, with only little overlap in the attendance at events. Over the years, the data has become a standard test case of network analysis algorithms, a meta-analysis of which can be found in~\cite{FreemanDavis}. We are interested in whether an inference based approach can assert the presence of latent classes and whether the class assignments found correspond to those suggested by the field experts.

First, we learn the standard stochastic block model with two latent classes for both actors and events, \ie model (\ref{ERGM2}), and only estimate class membership $\sigma_i$, $\tau_\mu$ and preference matrix $B_{rs}$. Figure \ref{DavisMatrix}a shows the data, with rows and columns of the attendance matrix reordered such that events/actors predominantly assigned to the same class are adjacent. The resulting block model is in contrast to findings of the original authors \cite{DeepSouth}. Events seem divided according to number of participants, \ie popularity, while actors seem divided according to number of events participated in, \ie activity. The expert classification due to social class is not correctly captured when trying to model the network through group effects alone. The reason for this failure is that under model (\ref{ERGM2}), the degree distribution for members of the same latent class is assumed to be Poissonian. The expected degree is the same for each member of a class. The inset in figure \ref{DavisMatrix}a shows that this assumption cannot capture the observed degree distribution. %The stochastic block model alone does not provide a mean unbiased model for the degree distribution of this network.
Since the standard stochastic block model does not model node degree independently of class preference; variance in degree distributions of both actors and events confuses the inference of group membership.

In contrast, the inset in figure \ref{DavisMatrix}b shows the expected degree vs. the observed degree when including activity and popularity parameters in the model as in (\ref{ERGM1}) and allowing for two classes. Now, the observed degree distribution can be accounted for. The introduction of activity and popularity parameters has also dramatic effects on the latent classes inferred. Figure \ref{DavisMatrix}b again shows the attendance matrix, but this time, rows and columns are ordered as given in \cite{DeepSouth} and the authors' assignment to social class is indicated by dashed boxes. The experts' classification matches almost perfectly that inferred using model (\ref{ERGM1}). We can see that events such as $8$ and $9$  which are attended by most actors receive high $\beta$ values and thus have very little discriminative power. Also, actors who are very active and occasionally participate in events predominantly frequented by actors from the other group such as Mrs.\ N.\ F.\ can still be assigned with high probability to a class, despite conflicting evidence in their participation record. Using model (\ref{ERGM1}) effectively allows one to decouple the preference effects of a group of actors for a group of events from global effects that contribute to the variance in node connectivity.
%
% FIG 3-------
\begin{figure}
\includegraphics[width=17.8cm]{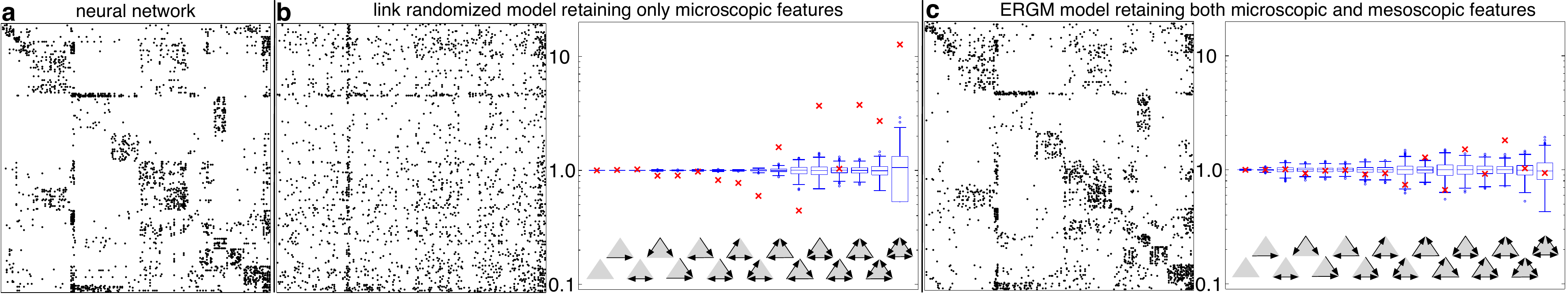}
\caption{Motif counts in the synapse network of \textit{C. elegans} compared to two random null models. \textbf{a)} Adjacency matrix of the observed neural network \cite{ChenCElegans}. \textbf{b)} Adjacency matrix of a typical realization of a link randomized version of the original data and resulting Z-score statistics of motif counts. Counts in the original data (red x) are compared to box plots of counts in 1000 link randomized null models. Strong deviations are found at 11 of the 16 motifs. Since the link randomized null models retain only node specific features, \ie the numbers of incoming, outgoing and reciprocated links at each node, the cannot capture the apparent mesoscopic structure in the original network and hence may over-estimate the statistical significance of some motifs. 
\textbf{c)} Adjacency matrix of a typical network generated from model (\ref{ERGM3}) with both node specific as well as class specific parameters estimated from the original network. 15 classes were used in this example. Using 1000 networks generated from model (\ref{ERGM3}) as a reference ensemble, the Z-score statistics show mild deviations only at 3 of the 16 motifs. This indicates that class structure may offer a more parsimonious explanation for the observed motif distribution.}
\label{Motifs}
\end{figure}
%----------

\textit{Caenorhabditis elegans:} In our second application, we explore the extent to which a dyadic model may explain the distribution of small sub-graphs, termed motifs, in a network. Motifs have received considerable attention as possible entities of network formation, \ie building blocks larger than single edges. Their distribution relative to random null models has been suggested to characterize entire classes of networks~\cite{AlonMotifs}. The over/under-representation of certain motifs with respect to random null models is often attributed to possible evolutionary pressures due to a motif's potential influence on the performance of the network's function~\cite{SuperFamilies,Reigl}.

We study the distribution of all $16$ possible $3$-node motifs in the $279$ neuron chemical synapse network of \textit{C. elegans}~\cite{ChenCElegans}. The null model commonly used to assess whether a particular motif is under- or over-represented in a network is generated by randomizing the original network conserving only microscopic structural features, \ie the number of incoming, outgoing and reciprocated links at each node is preserved. All other structural features and correlations are removed by the randomization. Figure~\ref{Motifs}b shows box-plots for motif counts in 1000 such random networks and in comparison the actual count of the 16 motifs in the chemical synapse network of \textit{C. elegans} normalized to the mean count found in the set of null models. We can see that using this null model, 11 of the 16 motifs are strongly over/under-represented and hence would qualify as possible starting points for further research on putative functional relevance.

% FIG 4-------
\begin{figure*}
\includegraphics[width=18cm]{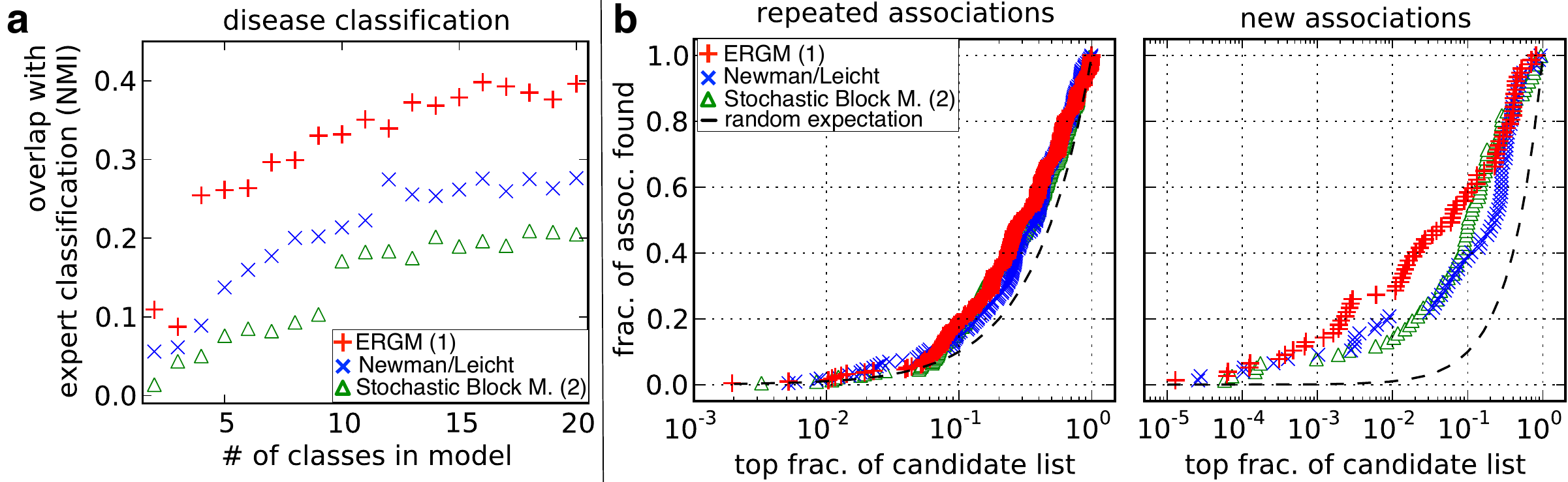}
\caption{Classification accuracy and predictive power of network models (\ref{ERGM1}), (\ref{ERGM2}) and that by Newman/Leicht (NL) \cite{LeichtMixture} . \textbf{a)} Overlap of an expert classification of diseases in the Diseasosome-Network \cite{Diseasosome} and that inferred using models and the data of known gene-disease associations recorded in the Online Mendelian Inheritance in Man (OMIM) database by Dec. 2005. Measure of overlap is normalized mutual information (NMI) \cite{FredJain}. \textbf{b)} Prediction accuracy at $16$ classes for confirmed associations added to the OMIM database between Dec. 2005 and Jun. 2010. For each model, a candidate list of associations is obtained by sorting all possible associations in descending order according to their probability under that model with parameters estimated from the Dec. 2005 data. We plot which fraction of actually confirmed associations is found in the corresponding top fraction of the candidate list. Entries due to new variants of a previously recorded association are listed as ``repeated associations'' while genuine new associations are reported as ``new associations''. For example: In the top $1\%$ of any candidate list, we expect to find $1\%$ of new associations due to chance alone. We do find $15\%$ of all confirmed new associations if the list was due to model (\ref{ERGM2}), $20\%$ if the list was due to the NL model and $30\%$ if the list due to model (\ref{ERGM1}).  See text for details.}
\label{OverlapPrediction}
\end{figure*}

%---------------

However, the standard null model also removes all mesoscopic structures, in particular structure due to groups of more than three nodes. A dyadic model such as (\ref{ERGM3}) lacks any parameter for three-node motifs but can generate an ensemble of null models that matches the observed network in terms of the observed node specific degrees as well as with respect to mesoscopic structural features. Such mesoscopic structure inevitably exists as neurons are located in different somatic regions and synaptic connections between closely located neurons are more likely than between distant ones~\cite{CommentonMilo}. Neurons are also aggregated in different ganglia making intra-ganglia connections more likely than inter-ganglia synapses. Furthermore, they serve different functions that influence their connectivity. For example, stimuli may be processed in a sensory neuron - interneuron - motor neuron cascade. The latent classes we infer from the data using (\ref{ERGM3}) can be explained using a combination of these factors (see suppl. material). For instance, some classes combine motor neurons only, some combine both sensory, inter- and motor-neurons which are all located in a particular somatic region, while others can be explained post-hoc as lying in the same ganglia. More important than the interpretation of these classes is whether a dyadic model that assumes all pairs of nodes as conditionally independent can account for the observed three node motif-counts in the network.

Figure~\ref{Motifs}c shows the box-plots of motif counts in 1000 networks generated from (\ref{ERGM3}) allowing for 15 different classes of neurons and using the parameters estimated from the original network, again normalized to the mean count. The comparison with the motif-count in the \textit{C. elegans} network now shows that only $3$ out of $16$ motifs cannot be explained by the null model. This result is remarkable as it underscores the importance of group specific effects in modeling complex networks. The fact that a simple dyadic model can explain a large portion of the three-node statistics in the observed data is a strong corroboration for our claim that latent classes of nodes are important determinants of network structure. Furthermore, it offers a very parsimonious explanation of motif statistics in this network and a more conservative estimation of their statistical significance.%highlights the importance the choice of null model has on determining which motifs, if any, are truly of functional importance.

%--
%\subsection{Online Mendelian Inheritance in Man}
\emph{OMIM:} As a last example, we study the influence of classification accuracy on the predictive power of probabilistic models using a bi-partite network known as the human ``Diseasosome-Network''~\cite{Diseasosome}. It represents known associations between genes and diseases recorded in the {OMIM} database~\cite{OMIM}. The network was first published in 2005 and we focus on the analysis of the largest connected component involving $516$ different diseases and $903$ different genes connected by $1550$ different associations known in 2005~\cite{Diseasosome}. The original publication provided an expert classification of the diseases into $22$ types. The type of disease is predominantly based on the tissues and organs involved (such as bone, connective tissue, muscular, dermatological, hematological, renal, etc.) or based on the affected system (such as skeletal, cardiovascular, immonological, metabolic or endochrinal, etc.)

To what extent does such a classification overlap with one inferred from a network of common genetic causes? We compare model (\ref{ERGM1}) with the less expressive standard stochastic block model (\ref{ERGM1}) and a more expressive model due to Newman and Leicht (NL)~\cite{LeichtMixture}. The latter includes both individual and group effects as in (\ref{ERGM1}), but instead of a single parameter for the overall activity or popularity of a node, it features one such parameter per latent class, \ie it models activity or popularity with respect to each class of nodes.

We compare the overlap between the expert classification of diseases and the one found algorithmically, based on the gene-disease association network alone.
%, measured as normalized mutual information (NMI)~\cite{FredJain} between the two classifications.
We restricted ourselves to using the same number of classes for both genes and diseases. The comparison of models (\ref{ERGM1}), NL and the standard stochastic block model (\ref{ERGM2}) is shown in figure~\ref{OverlapPrediction}a. As expected from the earlier discussion, neglecting individual node effects as in model (\ref{ERGM2}) reduces the overlap with an expert classification compared to model (\ref{ERGM1}). But, interestingly, the same applies when including gene-specific effects for every class of diseases and disease-specific effects for every class of genes as in the NL model. Too many explanatory variables per individual node seem to reduce the detection quality of latent classes.

Since 2005, the OMIM database has been steadily growing and $292$ new associations between those $516$ genes and $903$ diseases had been added until June 2010. Using the data from 2005 as a training set and these new additions as a test set, we compare the predictive power of the different models for future associations. New entries to OMIM comprise both new variants of already known gene-disease associations (repeated associations) as well as genuine new associations of genes with diseases that were not linked previously. Hence, the data offers the opportunity to differentiate predictive power with respect to these two types of entries. Using the parameters estimated from the 2005 data set for each model (\ref{ERGM1}), NL and (\ref{ERGM2}), we calculate the probability for association of each gene $i$ with each disease $\mu$ as $\prob{D_{i\mu}|\vec{\theta}}$. Then we sort these probabilities in descending order and hence obtain a candidate list for new or repeated associations. For instance, in the case of models with $16$ classes, figure~\ref{OverlapPrediction}B now shows how far we have to go down the candidate list to find a certain fraction of the associations that were added to the database over the course of 4 $1/2$ years.

Variants of already known associations seem to be added approximately randomly to the database as models (\ref{ERGM1}), NL and (\ref{ERGM2}) all perform close the random expectation for these repeated associations. For the genuinely new associations, however, we observe that all models strongly deviate from the random expectations. In particular (\ref{ERGM1}) outperforms both NL and (\ref{ERGM2}), with the latter two performing similarly.

Comparing figures \ref{OverlapPrediction}a and \ref{OverlapPrediction}b, we conclude that the standard stochastic block model may be too simple to capture the biologically relevant network structure as seen from a low overlap with an expert disease classification and low predictive power. With the inclusion of node specific effects, the NL model is more flexible in capturing the observed network leading to a higher overlap with the expert classification of diseases. It does, however, not generalize well as its predictive ability is still lower than that of the ERGM model (\ref{ERGM1}). The latter appears to provide the best compromise between flexibility and parsimony as it combines excellent ability to represent biologically relevant structures with high classification accuracy and a parsimonious inclusion of node-specific effects, leading to the best predictions.

%-------------------------
\section{Discussion}
We have presented an efficient, distributive algorithm that successfully estimates the parameters and latent group assignments of an exponential random graph model including both node specific and group specific properties. We have shown that including node specific effects in the estimation of latent classes leads to improved recovery of class assignments by domain experts. Additionally, we have shown that using a simple dyadic model, a large part of the triad statistics in networks may be explained, shedding new light on the discussion of motif distributions in complex networks. We expect our results to stimulate a discussion on the use of appropriate null models in the analysis of sub-graph distributions and their universality for certain classes of networks. Finally, we have explored the predictive power of the model to identify new gene-disease associations, using the OMIM database. Through these specific examples, we have demonstrated that node specific and group specific properties should be both incorporated when inferring and modeling structural features in complex networks.

%\begin{acknowledgments}
We would like to thank M. Weigt, S. Bornholdt, D.R. White, and J.P. Crutchfield for stimulating discussions. J.R. thanks the members of the Complexity Sciences Center at UC Davis for their hospitality.

This work was partially supported by the Volkswagen Foundation through a Fellowship Computational Sciences for J.R. and DAAD travel grants; support from EPSRC (EP/E049516) and the British Council ARC (1324) is acknowledged (D.S. and R.A.)
%\end{acknowledgments}

\end{small}

\begin{small}
\bibliography{../../../BibTex_Citations.bib}
\end{small}

%FIG 2 -------------

%------------------

% FIG 3-------
%----------

% FIG 4-------

\end{document}